# Unraveling Shadows: Exploring the Realm of Elite Cyber Spies

Fatemeh Khoda Parast*
khodapaf@uoguelph.ca
Faculty of Computer Science, University of Guelph, Canada

*Abstract*—The Equation Group, an advanced persistent threat identified by Kaspersky's Research Lab in 2015, was detected during the investigation of the Regin malware. Attributed to the United States National Security Agency, the Equation Group's techniques are more advanced than previously discovered threats. Despite being identified in 2015, detailed studies of their tactics, techniques, and procedures have been limited. This research examines the artifacts left by the group, revealing their advanced methodologies and analyzing the defensive mechanisms embedded within these artifacts designed to avoid detection by security systems. Additionally, solutions are proposed at various levels of the digital systems stack to counter the group's sophisticated attack strategies effectively.

## I. INTRODUCTION

Advanced Persistent Threats (APTs) are state-sponsored hackers using sophisticated techniques to carry out targeted attacks. These groups are often affiliated with government agencies and have access to extensive resources, including funding, personnel, and advanced tools [1]–[4]. During an investigation into the sophisticated malware Regin, Kaspersky uncovered an even more advanced malware. As the investigation continued, Kaspersky collected additional artifacts apparently developed by the same group. This APT group was named Equation Group (EQGRP) and attributed to the United States National Security Agency (NSA) based on the compelling evidence, such as activity timestamps matching US working hours and exploit methods similar to those used in US-attributed malware like Stuxnet [5], [6]. EQGRP employs various techniques, including web-based exploits, self-replicating worms, and the interception and manipulation of legitimate physical media and electronics. One notable method involves using USB sticks combined with various exploits to achieve its objectives [7]–[9].

One year after Kaspersky research lab discovered the EQGRP's malware binaries in the wild, a Russian-based APT group known as Shadow Brokers claimed to have stolen the EQGRP's hacking source codes. The leaked data contained two zip files: one protected by a password and the one publicly accessible. With the EQGRP's binaries in hand, Kaspersky precisely compared algorithm implementations, verifying the accuracy of Shadow Brokers' audacious claim. Shadow Brokers initially demanded $500 million in an auction to disclose the password for the source code. When the auction failed to attract buyers, they opted for a mass sale on Black Friday. Then, in April 2017, Shadow Brokers made a damaging decision by releasing the passwords. The outcome of this practice was numerous destructive attacks followed, leveraging techniques from the EQGRP code, including the infamous WannaCry and NotPetya ransomware attacks [10]. Figure 1 indicates the EQGRP artifact discovery timeline.

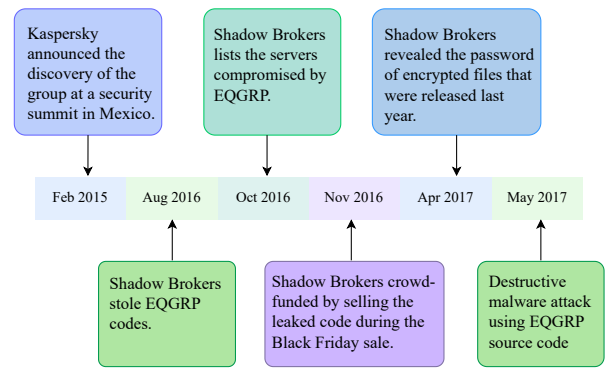

**Figure 1:** Equation Group Code Leake History

To date, EQGRP artifacts resemble a puzzle with many missing pieces. We have obtained numerous malware binaries designed for various stages of an attack, shared by Kaspersky—hosted by the famous malware sharing platforms, VirusShare[1] and Malware Bazzar[2]—and from a source code repository on GitHub titled

---

[1] https://virusshare.com
[2] https://bazaar.abuse.ch

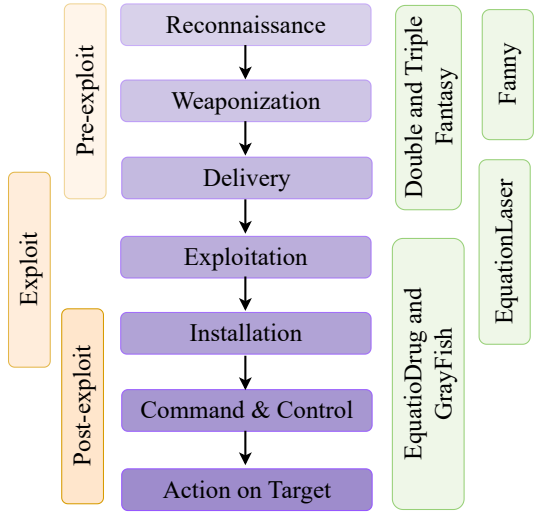

**Figure 2:** Exploit Platform Stages Corresponding to The Cyber Kill Chain Methodology.

*Lost in Translation*[3], which primarily focuses on the post-exploit stage. In this study, we connect the different parts of the puzzle and propose solutions to protect systems against the employed vulnerabilities.

This research focuses on the malware binaries caught in the wild to answer the following questions:

1) Define the attack target threat models.
2) Analyze how EQGRP infects targets through the exploit chain.
3) Investigate how attackers maintain persistence within compromised systems.
4) Examine the evasion techniques used by the group.
5) Develop strategies to protect systems throughout the attack chain using layered defence techniques.

The rest of this paper is organized as follows. Section II elaborates on the group's threat model. Section III discusses the group's techniques and attack platforms. Evasion techniques are discussed at Section IV followed by the protection and detection methods at Section V. Finally, we conclude and propose the future research direction on Section VI.

---
[3]https://github.com/x0rz/EQGRP_Lost_in_Translation

## II. THREAT MODEL

The Equation Group's victims span over 30 countries, including Iran, Afghanistan, Russia, Syria, the United States, Lebanon, Palestine, France, Germany, Singapore, Qatar, the United Kingdom, India, and Brazil. These victims come from a wide range of sectors, such as government and diplomatic institutions, telecommunications, aerospace, energy, nuclear research, oil and gas, military, nanotechnology, Islamic activists and scholars, mass media, transportation, financial institutions, and companies developing cryptographic technologies [11], [12].

Following the attack kill chain methodology, illustrated in Figure 2, attackers execute their objectives through various stages [13]. The EQGRP possesses a robust arsenal of tools to facilitate each attack stage. DOUBLEFANTASY and TRIPLEFANTASY serve to validate victims and ascertain their level of interest, maintaining a backdoor into the computer of potentially significant targets. EQUATIONLASER functions as a Trojan dropper, facilitating the initial infiltration of systems. EQUATIONDRUG and GRAYFISH serve as attack platforms, enabling the theft of information and facilitating the delivery of the objective on the target. FANNY operates as a worm designed explicitly for mapping air-gapped networks, further expanding the reach of the attackers [5], [6].

The EQGRP, with its specialized focus on air-gapped and non-air-gapped network intrusions, operates with a level of sophistication that demands our highest attention. The attack chain typically exploits zero-day vulnerabilities in non-air-gapped networks, whereas the physical access defines the primary model in the air-gapped systems, showcasing the group's advanced tactics [14], [15].

### A. Air-gapped Networks

Attacks using physical media, such as CD-ROMs, suggest using a highly stealthy tactic known as interdiction, where malicious actors intercept and replace shipped items with compromised versions [16], [17]. An incident that underscored the real-world implications of cyber attacks was the targeting of attendees of a scientific conference in Houston. Post-event, some participants received a seemingly harmless package by mail with conference proceedings and a slideshow of event materials. However, the enclosed CD maintained infected autorun software, a gateway to a potential system breach. The primary objective of the infected USB with FANNY malware was to map air-gapped networks using a unique USB-based command and control mechanism. Once the target was infected through the USB stick, FANNY identified the network components. The following stages of the attack varied depending on the objectives. If the goal were data collection, FANNY would gather essential system information, store it in the

USB's hidden section, and later transmit it when connected to an internet-enabled machine. Despite the limited hidden space, it could store critical information, including credentials. Attackers could encode instructions in the USB's hidden storage area to execute commands on air-gapped networks. When inserted into the air-gapped computer, FANNY would detect the system elements and execute the commands. This method allowed the EQGRP to control air-gapped networks via infected USB sticks while mapping their architecture [18].

*B. Non-air-gapped Network*

One of the group's techniques involves exploiting vulnerabilities in firewalls' Simple Network Management Protocol (SNMP) and Command Line Interface (CLI) [19]. SNMP is essential for managing network devices, enabling the exchange of management data among routers, switches, servers, and printers. Cisco firewalls offer administrators a CLI for configuring and managing the devices. Two models of Cisco firewalls, the Private Internet Exchange (PIX) and Adaptive Security Appliance (ASA), are particularly susceptible to these exploits. PIX is an earlier version of Cisco's standalone firewall technology, while ASA is its successor. ASA provides comprehensive support for firewalling, Virtual Private Networks (VPN), Intrusion Prevention Systems (IPS), Intrusion Detection Systems (IDS), and content filtering, making it a fully equipped security solution. Exploiting vulnerabilities in these systems involves more than unauthorized access; it includes crafting malicious commands or inputs to exploit parser weaknesses. This can result in severe consequences, such as executing unauthorized commands, bypassing access controls, modifying firewall configurations, or even gaining elevated privileges on the device [20], [21].

### III. TECHNICAL ANALYSIS OF EQGRP METHODS

The following conditions must be met to manipulate a non-air-gapped system through the SNMP vulnerability successfully: (1) SNMP must be enabled on the target system, (2) the community string must be known or discovered, (3) the attack must utilize IPv4, and (4) there must be access to systems on the activated firewall interface. The primary tools designed to exploit the system are EXTRABACON, EPICBANANA, and BANALRIDE.

EXTRABACON is implemented as a Python script designed to deactivate the firewall authentication. To deliver the exploit, the attacker must know the community string—which functions similarly to a password—and the target's IP address. Using the script and providing the necessary details, attackers can gather information about the target machine to deactivate password-based authentication (Listing 1, lines 3 and 4). Once EXTRABACON deactivates the authentication, EPICBANANA exploits a local command-line buffer overflow vulnerability to implant the next stage payload. EPICBANANA uses a precisely crafted input to execute arbitrary code on the target machine, potentially taking control of the entire system (Listing 1, line 8). Afterward, communication with the firewall via BANALRIDE becomes possible, enabling further interaction and exploitation of the system through established connections. Using the *bride* tool, attackers first communicate with the firewall, exchange a 21-byte UDP packet to establish a Diffie-Hellman key, and create an encrypted channel (Listing 1, line 12). Finally, they install BANAGLEE into the firewall's memory. This established channel facilitates data transfer and execution of further commands, expanding the attacker's control and manipulation over the compromised system. Table I summarizes the pre-exploit chain artifacts and indicates the corresponding Common Vulnerabilities and Exposure (CVE) values. The attack scripts are available within the *Lost in Translation* GitHub repository.

```
1  // EXTRABACON
2
3  $ ./extrabacon.py target_machine_info -t <victim_ip> -c
       password
4  $ ./extrabacon.py exec -k target_machine_info -t
       <victim_ip> -c password --mode pass-disabled
5
6  // EPICBANANA
7
8  $ ./epicbanana.py  -t <victim_ip>   --proto=ssh
       --username=user_name --password=the_pass
       --arget_vers=asa804   --mem=NA -p 22
9
10 //BANALRIDE
11
12 $ ./bride-1100 --lp <victim_ip> --implant  <attacker_ip>
       --sport <victim_port> --dport <attacker_port>
```

**Listing 1:** Exploit Chain [22]

*A. Exploit Platforms*

EQUATIONDRUG and GRAYFISH are the primary cyber espionage platforms used by the group, both utilizing boot manipulation techniques. In a typical boot process, the CPU starts in real mode and executes the Basic Input/Output System (BIOS). The BIOS then identifies and transfers control to the Master Boot Record (MBR). The MBR, located in the first sector of the system's hard disk, navigates the partition table (PT) to locate the designated bootable partition, where the operating system (OS) is usually stored. Control is then handed over to the Volume Boot Record (VBR) within the bootable partition. Once the MBR has determined the bootable partition, the VBR code loads the first stage of the bootloader into memory and passes control to it. The bootloader retrieves additional code from the disk, switches to protected mode, and then loads and executes the kernel [23], [24]. Figure 3 illustrates the boot sequence in real and protected mode.

Table I EQGRP Exploit Chain Tools.

| Exploit | CVE | Vulnerability | Description |
| --- | --- | --- | --- |
| EXTRABACON | CVE-2016-6366 | Turn the firewall password ON or OFF | Remote code execution |
| EPICBANANA | CVE-2016-6367 | Exploit SSH or Telnet | CLI remote code execution |
| BANALRIDE | N/A | N/A | Establish secure channel with C&C server, load BANAGLEE |
| BANAGLEE | N/A | N/A | Communicate with C&C server, network traffic manipulation |
| JETPLOW | N/A | N/A | Persistent BANAGLEE between reboots |

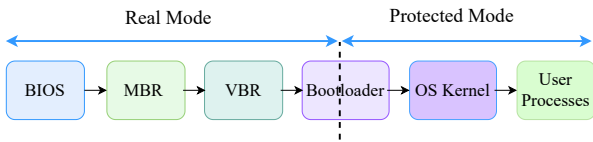

**Figure 3:** BIOS-based Systems Boot Procedure [23]

Within the disk layout, specific segments are designated as dark regions, which exist outside the boundaries of a filesystem. Since these regions are not part of any filesystem structure, they remain hidden and inaccessible during regular system operations. Examples of dark areas include the MBR, VBR, bootloader, inter-partition gaps, and the space extending beyond the last partition to the physical end of the disk. These gaps often span several megabytes on most systems. Dark regions appeal to attackers because they are infrequently modified, typically only during major OS updates or other rare events. Additionally, most protection mechanisms overlook these regions since they primarily focus on the filesystem level [23].

EQUATIONDRUG is an intricate exploit platform that outperforms a Trojan to function as a comprehensive espionage tool, facilitating cyber espionage by deploying tailored modules onto targeted victims' machines. At its core, EQUATIONDRUG comprises a collection of drivers, a central orchestrator, and various plugins, each with a unique ID and version number defining its specific functions. The platform includes default plugins for tasks such as file retrieval and screenshot capture. It enhances its sophistication by encrypting stolen data within a proprietary virtual file system before transmitting it to C&C servers. The platform initiates with the kernel mode driver component, which waits for system startup and then triggers the execution of the user-mode loader *mscfg32.exe*, subsequently launching the central orchestrator module from *mscfg32.dll*. Both intrinsic and supplementary components may load additional drivers and libraries as needed [2], [5], [18].

GRAYFISH, as the upgraded version of the EQUATIONDRUG, employs a bootkit for persistency, embedding its code into the MBR during computer startup, thereby taking control of the OS's loading process—the platform stores stolen files in an encrypted Virtual File System (VFS) to secure its presence. The malware's loader employs a robust encryption method, utilizing SHA-256 iterated one thousand times over the unique NTFS object ID of the victim's Windows folder, ensuring that decryption requires this specific information. This complex encryption scheme enhances security by making reverse engineering difficult and decryption nearly impossible without the exact NTFS object ID. Additionally, GRAYFISH uses evasive tactics by self-destructing entirely in case of startup errors, leaving no traces behind [25], [26].

GRAYFISH utilizes a Windows kernel rootkit to conduct its malicious operations in the highly privileged Ring 0 mode (kernel mode). This rootkit facilitates code injection into running processes and employs several tricks to confuse security researchers analyzing live systems for kernel-mode anomalies. During the DriverObject initialization, the rootkit retrieves the hardcoded process ID within its code and attempts to acquire the address of its process object using PsLookupProcessByProcessId. Subsequently, it creates a buffer or data storage area to store information about specific kernel functions of the targeted OS. This buffer is organized with each entry consisting of two parts: the hash value of the function's name and the function's address. The hash value, serving as a unique identifier derived from the function's name, acts as shorthand, while the address indicates the function's location within the kernel's memory space. Figure 4 illustrates the exploit chain in a high-abstract manner where the attacker starts by looking up the process ID to highjack the kernel driver.

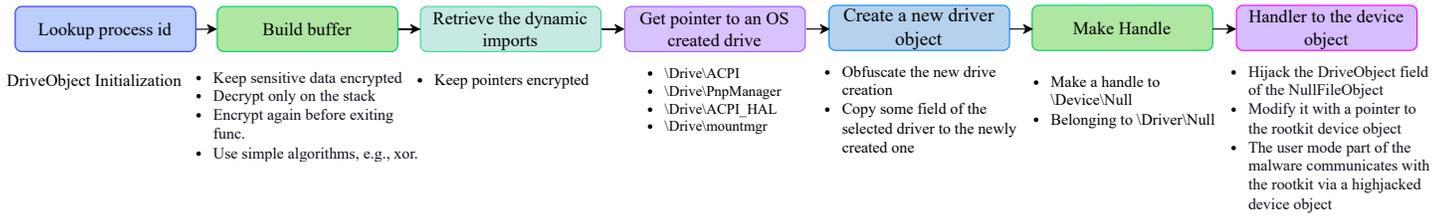

**Figure 4:** Rootkit OS Kernel Driver Highjack

Hard drives, a crucial component of any system, have a controller with firmware stored in a memory chip or flash ROM. When EQUATIONDRUG or GRAYFISH infiltrates a system, the firmware flasher module connects to a command server to download and implant a malicious payload code into the firmware, replacing the legitimate version. This trojanized firmware, a proof to the malware's persistence, allows attackers to maintain access, even after software updates or OS reinstalls, as it can restore wiped components by contacting the command server. It's important to note that vendor firmware updates may not remove the malicious code, as updates often replace only firmware parts.

## IV. EVASION TECHNIQUES

The EQGRP applies various evasion techniques in different stages of an attack, making intrusion detection a complex procedure. This section discusses the most prevalent evasion techniques used through the group's exploit chain.

**Fileless Malware:** This method avoids the need to write any files to disk, thus reducing the chances of detection. Once the initial access is gained, the EQGRP's tools, such as EQUATIONDRUG and GRAYFISH, deploy payloads that reside entirely in memory. These tools can load additional modules and execute commands directly within the memory space of compromised systems [27].

**Anti-Debugging and Anti-Reversing Techniques:** Malicious code often employs anti-debugging and anti-reversing techniques to prevent security researchers from analyzing it. These techniques include obfuscating code, detecting debuggers, and employing anti-analysis checks to make dynamic and static analysis more difficult. The groups' malware binaries are packed, making static analysis almost useless. String extraction, performed to extract suspicious strings, such as IP addresses and domain names, is not applicable unless the malware is executed [28].

**Anti-Virtualization Techniques:** Malware may include anti-virtualization techniques to detect if it is running within a virtualized environment, such as a virtual machine or sandbox, and alter its behaviour to avoid detection or analysis. The method can involve checking for virtualized hardware, registry keys, or specific virtualization artifacts [29]. The group's malware includes calling *SleepEx* function to halt the execution. The function calculates the delay period for execution, assuming that the sandbox session will terminate before this delay period elapses.

**Staged Payloads:** The group employs staged payloads, where the initial payload is benign or harmless, and only after some time or specific conditions are met does it download and execute the actual malicious payload. This method evades detection by delaying the execution of malicious activities until after initial security checks have passed [30]. The EQGRP uses a multi-stage payload where, in case of any error, the malware has the self-destruction ability to remove any traces in the system. Only if a stage is successful will the next stage payload be transferred to the target.

**Traffic and Communication Obfuscation:** Malicious actors obfuscate the network traffic and communication channels using encryption, tunnelling, or legitimate protocols to blend in with regular network traffic. Obfuscation makes it harder for network-based detection systems to identify and block malicious activity [31]. The EQGRP extensively uses encryption at every stage, including the malware code, the stolen data storage within the VFS, and communications with the C&C server.

## V. PROTECTION AND DETECTION MECHANISM

The Equation Group targets air-gapped and non-air-gapped networks, making comprehensive security measures essential for organizations. A multifaceted approach is necessary to secure an organization's ecosystem effectively. While deploying advanced solutions such as firewalls, VPNs, IDS, and IPS is critical to prevent unauthorized access, these measures alone are insufficient [32]. Regular software patching, physical access control enforcement, and user awareness are equally vital steps to reduce the likelihood of successful attacks. Beyond these standard measures, specialized protection mechanisms are required to address the specific vulnerabilities the EQGRP utilizes to exploit the systems.

**Safe listed devices:** A security detection and prevention method is used to enhance the protection of a network or system by allowing only approved, trusted devices to access the network or system resources. Creating and maintaining a safelist can be resource-intensive, requiring ongoing management and updates. If a legitimate device is not correctly safe listed, it could be denied access, potentially disrupting business operations. To address the challenges, implementing a two-factor authentication system for USB storage devices can be applied to protect systems from malicious activities carried out through USB devices [33].

**VLAN (Virtual Local Area Network) segmentation** VLAN segmentation involves dividing a physical network into multiple distinct logical networks. Each VLAN behaves like an independent network, even though multiple VLANs may share the same physical infrastructure. Configuring VLANs can be complex, especially in large networks with many devices. Maintaining and updating VLAN configurations can become increasingly complicated as the network evolves. Inter-VLAN routing, necessary for enabling communication between different VLANs, adds another layer of complexity [34].

**Overflow preventive:** Buffer overflow vulnerabilities occur when a program writes more data to a buffer than it can handle, potentially leading to memory corruption and exploitation by attackers. To prevent buffer overflow attacks, developers can implement various preventive measures, such as bounds checking, input validation, stack canaries, and address space layout randomization (ASLR). Methods such as stack canaries, bounds checking, and runtime protections can introduce significant performance overhead. Older software might not be compatible with new security features or might require significant modifications [35], [36].

**Firmware manipulation protection:** The NIST Special Publication 800-147 outlines a comprehensive solution for securing firmware through several vital guidelines. It mandates using an authenticated BIOS update mechanism, employing digital signatures to ensure only authenticated updates are applied. An optional secure local update mechanism allows updates with physical presence, ensuring authenticity and integrity. Integrity protection mechanisms must prevent unauthorized BIOS modifications, typically enforced by hardware protections. Additionally, the update process must be non-bypassable, ensuring no component can override the authenticated update mechanism, safeguarding against unauthorized BIOS modifications [37], [38].

## VI. CONCLUSION AND FUTURE WORK

This research is a brief study of the Equation Group's puzzle. The investigation reveals a highly sophisticated and resourceful cyber-espionage entity attributed to the United States National Security Agency. This group employs advanced techniques to infiltrate air-gapped and non-air-gapped networks, leveraging zero-day vulnerabilities and innovative evasion methods. The analysis of their tactics, techniques, and procedures highlights their ability to remain undetected for extended periods, compromising numerous sectors across the globe. The Equation Group's use of tools like EQUATIONDRUG and GRAYFISH, which exploit deep system vulnerabilities, underscores the need for robust, multi-layered defensive strategies. The impact of their operations, particularly in the wake of the Shadow Brokers' leaks, demonstrates the potential for significant global disruptions and the ongoing challenge of defending against such advanced persistent threats.

Future research should focus on developing more advanced detection and prevention mechanisms tailored to the unique challenges posed by groups like the Equation Group, including enhancing real-time monitoring capabilities, employing machine learning algorithms for anomaly detection, and improving the resilience of firmware against sophisticated attacks. Additionally, a collaboration between government agencies, private sector companies, and international cybersecurity organizations is essential to share threat intelligence and develop unified defence strategies. Further studies should also explore the socio-political implications of state-sponsored cyber espionage, aiming to establish more evident norms and regulations to mitigate the risks associated with such activities. While this study analyzed the pre-exploit and exploit stages using binary artifacts, future studies can focus on post-exploit tools. These tools are invaluable assets for security professionals and offer significant potential for research in security monitoring and management tasks.